\pdfoutput=1
\documentclass[12pt,authoryear]{imsart} 

\usepackage[T1]{fontenc}

\usepackage[dvipsnames]{xcolor}

\usepackage{amsmath,amssymb}

\usepackage{graphicx}
\usepackage{placeins}   
\usepackage{float}

\usepackage[round,authoryear]{natbib}
\let\textcite\citet
\let\parencite\citep
\let\autocite\citep
\usepackage[hidelinks]{hyperref}

\usepackage{microtype}

\begin{document}

\begin{frontmatter}

\title{Forecasting Arctic Temperatures With Quantile Machine Learning}
\runtitle{Melting Days in Svalbard}

\begin{aug}
  \author{\fnms{Richard} \snm{Berk}}
  \address{University of Pennsylvania\\
           USA\\
           Corresponding Author: \texttt{berkr@sas.upenn.edu}\\
           ORCiD: \url{https://orcid.org/0000-0002-2983-1276}}
  \runauthor{R. Berk}
\end{aug}

\begin{abstract}
Using data from the Longyearbyen weather station, quantile gradient boosting (``small AI'') is applied to forecast daily temperatures in Svalbard, Norway. Temperatures above $0^{\circ}\mathrm{C}$ are of special interest because of their impact on ice, snow, and tundra permafrost. To improve forecasting skill for warmer temperatures, the target quantile is 0.60; forecast underestimates are weighted 1.5 times more heavily than forecast overestimates when the quantile loss is computed. Predictors include eight routinely collected indicators of weather conditions, each lagged by 14~days, yielding temperature forecasts with a two-week lead time. Adaptive conformal prediction regions quantify forecasting uncertainty with provably valid coverage. Using a holdout sample, a forecast of $> 0^{\circ}\mathrm{C}$ is correct 14~days later at least 80\% of the time. Implications for Arctic adaptation policy are discussed.
\end{abstract}


\end{frontmatter}

\section{Introduction}

The oceans and cryosphere interact to support unique ecosystems while exchanging water, energy, and carbon with Earth's climate system. The 2019 \textit{Intergovernmental Panel on Climate Change} (IPCC) report concludes that global warming has altered this interaction, causing ``mass loss from ice sheets and glaciers, reductions in snow cover, decreases in Arctic sea-ice extent and thickness, and increased permafrost temperatures'' \citep[A.1]{IPCC2019}. Ecosystem impacts have been equally dramatic \citep[A.4]{IPCC2019}. 

Derived from computer simulations such the \emph{Community Earth System Model}, these are macro-scale conclusions for spatial units from about $10^5$ meters to about $10^8$ meters. There is a need as well for spatial units at local-scales of $10^2$ meters or less \citep{Oke1987, Danabasoglu2020, CESM2Grids}. Downscaling methods have been proposed, but they rest on strong assumptions about the parent Global Climate Model, and no consensus has emerged on a preferred approach \citep{Nishizawa2018}. 

The most recent numerical weather prediction methods such as NOAA's \emph{Research and Forecasting Model} can provide output at local scales \citep{ECMWF2023, NOAAGFS2023, NOAAHRRR2023}, but the grid overlaid takes no account of topographical features or political boundaries. A grid cell might span a coastal mountain range and the Pacific Ocean, averaging over two very different climates. It might also straddle political boundaries — parts of several counties, or portions of two countries — each with different authorities, budgets, and policy instruments. A forecast for such a cell is not merely imprecise; it is operationally useless, because no single decision-making authority has jurisdiction over it.

Estimating forecasting uncertainty also can be problematic as well. \textcite[12]{Gettleman2016} summarize the challenges for climate models that apply equally to weather forecasting: ``Uncertainty in climate models has several components. They are related to the model itself, to the initial conditions of the model, and to the inputs that affect the model. All three must be addressed for the model to be useful.'' Despite many years of trying, obtaining valid estimates of uncertainty remains problematic \citep{Fu2025}.

There have been recent advances in climate and weather forecasting \citep{Pathak2022, Price2024, Bodnar2025}. Yet issues of coarse spatial scales, and equivocal uncertainty estimates remain largely unresolved \citep{Ortega2022, Balaji2022, Chang2023, Zhang2025}. In this paper, an approach is offered, as a complement to existing climate and weather forecasting, that aims to make some progress on both matters. 

Svalbard, Norway, is used as a testbed for which forecasts of Arctic temperatures are made. Temperatures are the focus because of legitimate concerns about a warming climate. The Svalbard data can resolve spatial detail precisely where the forecasts are most needed. Daily forecasts are constructed using quantile supervised learning, that can be seen as a a form of ``small AI.''  There also are valid measures of forecasting uncertainty well suited for highly local and rapid adaptations to Arctic warming.

Section~2 briefly provides some background on Svalbard. Section~3 describes the data, the construction of holdout samples, and the 14-day lagging of predictors used to support legitimate forecasting. Section~4 introduces the statistical methods, focusing on quantile supervised learning. Section~5 presents the empirical results, emphasizing visualization and algorithm interpretations. Section~6 addresses uncertainty through adaptive conformal prediction regions for multiple time-series data and offers a grounded way to communicate forecast reliability to stakeholders. Section~7 discusses policy implications, with emphasis on short-fuse adaptations to Arctic melting, and Section~8 concludes. A synopsis of the data analysis steps is provided in Appendix~A.

\section{Svalbard, Norway: The Forecasting Setting}

Svalbard is a remote Norwegian archipelago in the Arctic Ocean, located about halfway between mainland Norway and the North Pole. Its main settlement, Longyearbyen, lies at roughly $78^{\circ}\mathrm{N}$ latitude, well above the Arctic Circle. Only a small fraction of the land is vegetated, mostly tundra, while the remainder is dominated by permafrost, ice, and bare rock. Longyearbyen is one of the northernmost permanently inhabited places on Earth.

The maritime climate is shaped substantially by Atlantic Ocean currents. Average winter temperatures in Longyearbyen range from about $-20^{\circ}\mathrm{C}$ to $-14^{\circ}\mathrm{C}$, while summers are typically between $3^{\circ}\mathrm{C}$ and $7^{\circ}\mathrm{C}$.  Temperatures approaching $0^{\circ}\mathrm{C}$ cause melting of sea ice, glaciers, ice sheets and snow cover.  Local infrastructure can be threatened in ways that require adaptations of many kinds. Melting permafrost has ecological consequences for plant and animal life. In recent years, summer temperatures have more often spiked above $10^{\circ}\mathrm{C}$.

\section{Data}

The data come from the Longyearbyen weather station at the airport in Svalbard, and are easily downloaded from the Integrated Surface Database using the \emph{worldmet} package in \textsf{R}. The Database contains weather station data from around the globe. It uses the same standard format regardless of origin. The Longyearbyen weather station is meant to inform airport operations and characterize meteorological condition where the vast majority of people live. It less useful beyond a few kilometers outside of Longyearbyen because the terrain changes significantly.

The downloaded data are for the full years of 2022, 2023, and 2024. The response variable is the daily, solar-time 2~p.m.\ air temperature in degrees Celsius at the Longyearbyen weather station. Each daily temperature value at 2~p.m.\ solar time is an \emph{instantaneous observation}—a snapshot of conditions with a constant daily reference time rather than an hourly or daily mean. The 2~p.m. solar-time convention is used uniformly throughout the year to provide a consistent temporal reference for forecasting. It is a good proxy for the highest daily temperature.

The relevant weather station predictors  include (1) wind direction in degrees from true north, (2) wind speed in meters per second, (3) air temperature in degrees Celsius, (4) atmospheric pressure in hectopascals (hPa), (5) visibility in meters, (6) dew point in degrees Celsius, (7) relative humidity in percent, and (8) a day counter ranging from~1 to~365. The counter can capture temporal trends: on average, the diurnal months are warmer than the nocturnal months, although temperature changes over time may be nonlinear. Several predictors are likely to interact in complex ways \citep{Semenov2021}.

Predictors are lagged by 14 days, a choice consistent with some recent climate research \citep{Li2024} and one that provides operationally useful advance warning for local decision-makers. The 14-day lags translate into fitted 2~p.m.\ temperatures two weeks later that are a foundation for forecasts 14~days in advance.\footnote
{
Lagging variables is a routine procedure in feature construction. In this case, however, lagging the first 14~days of a year pushes their reference points into the final 14~days of the previous year. Data from 2021 are included to provide lagged values for the first 14~days of 2022. Otherwise, the 2021 data are not used.
}

Anticipating the use of adaptive conformal prediction regions to properly address forecast uncertainty \citep{Romano2019}, a variant on split-sample methods is employed. Observations for 2023 constitute the training data. Calibration data from 2022 can provide, under exchangeability, the basis for valid nonconformal scores. The calibration data are divided into observations for which the forecasted temperature is $> 0^{\circ}\mathrm{C}$ and observations for which the forecasted temperature is $\le 0^{\circ}\mathrm{C}$. The melting of snow, sea ice, glacier ice, and permafrost creates positive feedback loops that change the manner in which temperature variation is produced. On subject-matter and statistical grounds, adaptive conformal methods are therefore applied separately to each subset.\footnote
{
One might wonder why the data splits are made using forecasted temperatures rather than observed temperatures. Observed temperatures are certainly available in the weather station data. But when real forecasting is, only the forecast is known. If a future temperature were known, there would be no need to forecast it.
}

Data from 2024 serve as a pristine holdout sample treated as ``new cases'' to document true forecasting skill. As test data, these observations have no role whatsoever in training. The data split by forecasted temperature applied to the calibration data is applied again. The requirements necessary to capitalize properly on all three datasets within the split-sample approach are discussed in Section~6. 

The complete dataset forms a multiple time series, making temporal dependence a potential complication. Split samples are commonly disjoint subsets created by \emph{random sampling} from the data available, but that is likely to obscure any temporal dependence \citep[sec.~5.8]{Hyndman2021}. A key requirement is that the same physical processes apply during the identical months of 2022, 2023, and 2024; data for all three years are treated as random realizations from the same underlying joint probability distribution. Nonetheless, the pace of Arctic amplification may affect this comparability, an issue examined empirically in Section~5.2.

\section{Statistical Methods}

The analyses to follow produce legitimate forecasts from supervised machine learning applied to temporal data. Holdout samples are used for forecasting and to construct provably valid prediction regions. The statistical framework is an interlocking set of procedures. To the author's  knowledge, this combination of methods has not previously been applied to Arctic temperature forecasting. In the next section, each component is discussed in a grounded manner as the procedures are applied.  A step-by-step statistical synopsis is provided in the appendix.

\subsection{Some Details}

The Arctic multiple time-series data described above are analyzed using quantile gradient boosting, with the 60th percentile (i.e.,~$Q(0.60)$) as the estimation target. Let $y$ denote the numeric response variable, $\hat{y}$ its fitted value, and $\tau$ the target conditional quantile. Quantile gradient boosting minimizes the following loss function \parencite{Koenker1978}:

\begin{equation}
L_{\tau}(y, \hat{y}) =
\begin{cases}
\tau \cdot (y - \hat{y}), & \text{if } y \ge \hat{y}, \\[4pt]
(1 - \tau) \cdot (\hat{y} - y), & \text{if } y < \hat{y}.
\end{cases}
\end{equation}

With $\tau = 0.60$, underestimates receive $0.6/0.4 = 1.5$ times more weight than overestimates when the loss is computed. For this application, about 40\% of the 2023 Longyearbyen temperatures exceed $0^{\circ}\mathrm{C}$. Because relatively high values tend to be underestimated and relatively low values tend to be overestimated, fitting the conditional 0.60 quantile forces the gradient-boosting algorithm to work harder to avoid underestimates. This occurs most often among the warmest 40\% of temperatures. Indirectly, therefore, these ``melting days'' are weighted more heavily than the rest. Melting days are a particular concern because practical climate adaptations at a local level are often needed.

Adaptive conformal prediction regions are used to obtain, even in finite samples, valid measures of forecast uncertainty. They are adaptive because they capture each forecast's particular amount of uncertainty. The 2022 calibration data are used with the results the quantile gradient boosting algorithm to obtain residuals that serve as provisional nonconformal scores. They are provisional because their desirable properties depend on the exchangeability; they are only defensible as proper nonconformal scores when exchangeability is not violated for the residuals.\footnote
{
A dataset is exchangeable when the order in the observations are realized does not affect their probability distribution. IID data are exchangeable data, but exchangeable data are not necessarily IID.
}
For the weather station data, temporal dependence in the calibration data residuals, which violates exchangeability, is removed with time series procedures \citep{Box2015}.

The forecast data from 2024 are used to obtain true forecasts based on the results of the quantile gradient boosting algorithm. They are true forecasts because the 2024 data are pristine. They are not used in the training or in the construction of the nonconformal scores. In real forecasting applications, the predictor values are newly realized and therefore unavailable during model training — the same condition that holds here.

\section{Results}

The results begin with simple univariate plots to provide context for the more involved analyses. Because of the months-long alternation between daylight and darkness, Arctic temperatures fluctuate differently from those at lower latitudes. The figures that follow should be largely self-explanatory.

\subsection{Univariate Plots}

\begin{figure}[htbp]
\centering
\includegraphics[width=0.90\linewidth]{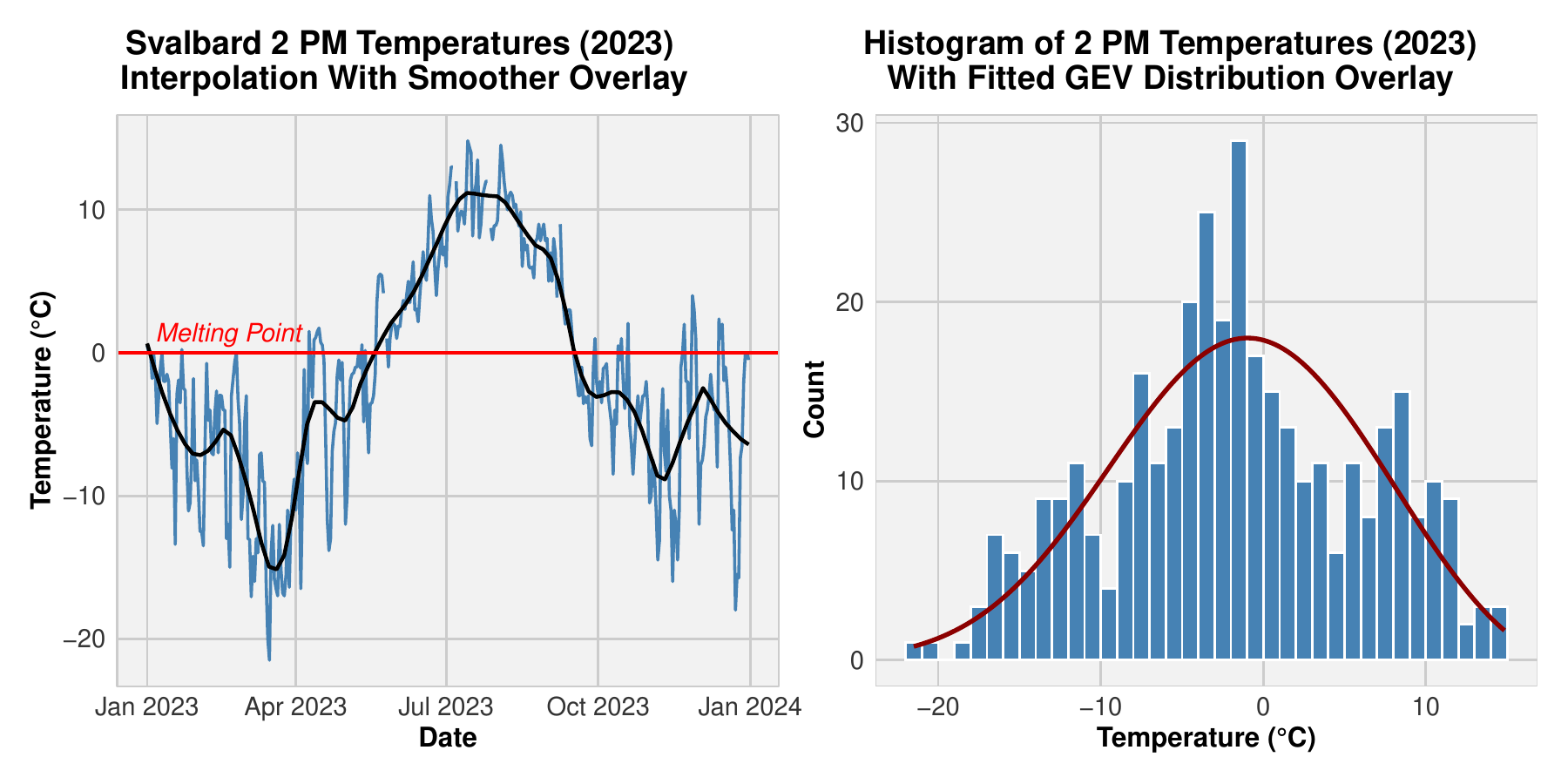}
\caption{2~p.m.\ 2023 temperatures with the time-series plot in the left panel and the histogram with a GEV distribution overlay in the right panel.}
\label{fig:univariate}
\end{figure}

A time series of the 2~p.m.\ temperatures for 2023 should reflect the expected Arctic seasonal swings. Figure~\ref{fig:univariate} (left panel) shows precisely that. The light-blue irregular line interpolates the daily temperatures, while the smooth black line shows the results of a loess smoother applied to those values. The red horizontal line marks the melting point at $0^{\circ}\mathrm{C}$. Temperatures during the nocturnal months are, on average, colder than during the diurnal months, and transitions between them are gradual. Melting temperatures are common from early June through late September.

The right panel displays a histogram of the 2~p.m.\ temperatures for 2023 with a generalized extreme value (GEV) distribution overlay. Temperatures range from slightly below $-20^{\circ}\mathrm{C}$ to slightly above $10^{\circ}\mathrm{C}$, revealing substantial variability. Although the GEV distribution is frequently applied in studies of high temperatures \citep{McKinnon2022}, it adds little interpretive value here.
The histogram is roughly symmetric, although the left tail is somewhat longer than the right. No clear outliers are apparent. The extreme temperatures in both tails are essentially a continuation of the overall distribution. They are not otherwise distinctive.

\subsection{Time-Series Plot for Svalbard Temperatures in 2022, 2023, and 2024}

Recall the premise that the 2022 calibration data, the 2023 training data, and the 2024 forecasting data represent realizations from the same underlying joint probability distribution; the underlying physics for corresponding months in those three years should be comparable. However, there could be aberrations perhaps caused by rapid Arctic amplification.

\begin{figure}[htbp]
\centering
\includegraphics[width=0.75\linewidth]{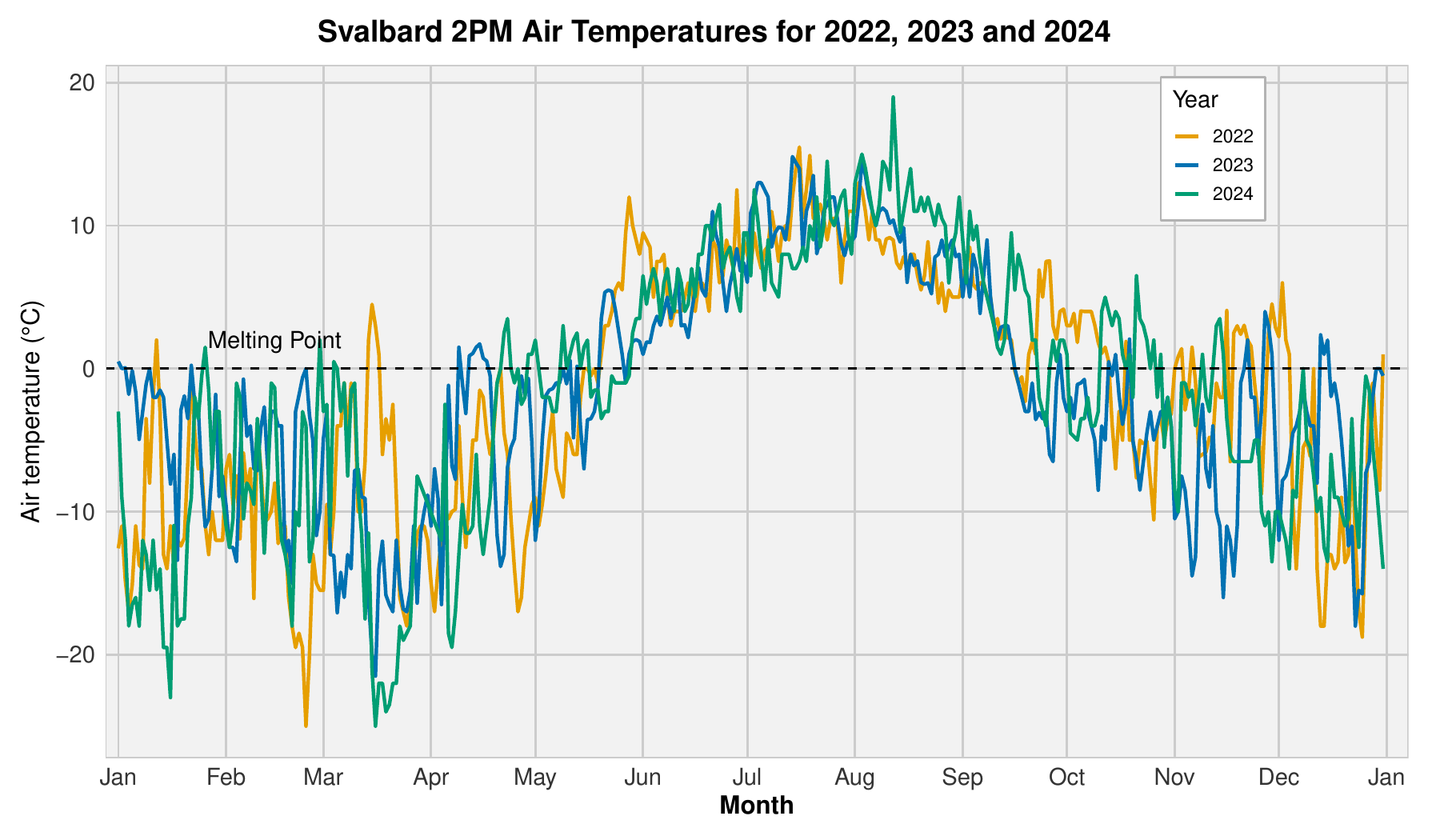}
\caption{Daily 2~p.m.\ temperatures for 2022, 2023, and 2024 using a color palette that is color-blind friendly.}
\label{fig:multiple}
\end{figure}

Figure~\ref{fig:multiple} provides a visual comparison of the three temperature time series. The jagged lines show the 2~p.m.\ daily temperatures plotted against day of year. The dashed horizontal line marks the melting temperature at $0^{\circ}\mathrm{C}$.

Figure~\ref{fig:multiple} shows substantial overlap across the three years. Seasonal trends are captured similarly, both in temperature values and in timing. The three time series move largely in lockstep from early to late summer, during which melting temperatures are nearly universal. There appears to be no compelling reason to reject the claim that each series consists of random variables drawn from the same joint probability distribution. In practice, the 2022 and 2024 temperatures serve as promising holdout samples for evaluating forecasts of the 2023 temperatures.

\subsection{The Quantile GBM Fit of Temperature}

Some changes in the predictor variables were required. Visibility and atmospheric pressure were dropped because of excessive amounts of missing data. Wind direction (reported in degrees) was transformed into its sine and cosine components after converting degrees to radians. This representation captures the circular nature of wind direction, for which 0° and 360° denote the same physical direction.\footnote
{
Wind direction was recorded in degrees but is intrinsically circular, with 
$0^\circ$ and $360^\circ$ representing the same direction. Each direction 
$\theta$ was therefore converted to radians and encoded as $\sin(\theta)$ 
and $\cos(\theta)$. Using both components is necessary because neither 
$\sin(\theta)$ nor $\cos(\theta)$ alone uniquely identifies a direction 
(e.g., $\sin(30^\circ)=\sin(150^\circ)$), whereas the pair 
$(\cos\theta,\sin\theta)$ provides a unique and smooth representation of 
all possible directions on the unit circle.
}

With the quantile parameter fixed at $\tau = 0.60$, the quantile gradient-boosting procedure in~R ran efficiently. The shrinkage value was set to 0.0001, the interaction depth to~6, and the minimum number of observations in a terminal node to~6. These values were chosen to foster slow convergence so that the right tail of the temperature distribution would be fitted more accurately. Using 2021 data as a holdout sample \footnote
{
Recall that the 2021 data were used to obtain lagged values for the predictors during the earliest 2 weeks of the training data. The construction these lagged values precedes the training. 
} 
to protect against overfitting indicated that approximately 27{,}000 iterations were appropriate for these data.\footnote
{
With a substantially larger shrinkage value, many fewer iterations might have been adequate. However, the best shrinkage value could not be known in advance, and empirical tuning encourages ``cherry picking'' that can undermine later statistical inference \parencite{Kuchibhotla2022}. 
}

\begin{figure}[htbp]
\centering
\includegraphics[width=0.75\linewidth]{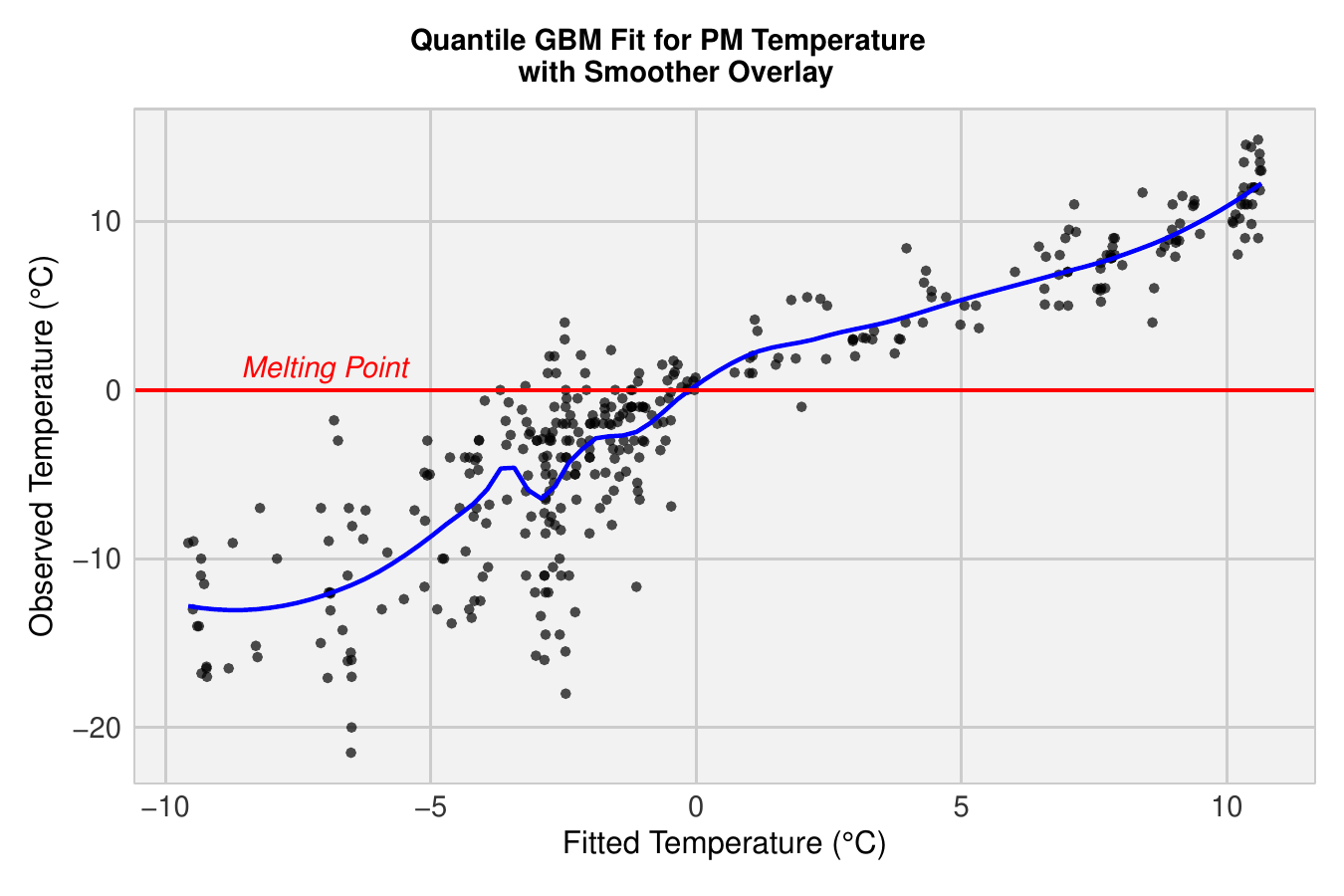}
\caption{Observed versus fitted 2~p.m.\ temperatures. Black dots are the data, the blue solid line is a loess smooth serving as a visual aid, and the red horizontal line marks the melting point at $0^{\circ}\mathrm{C}$.}
\label{fig:fit}
\end{figure}

Figure~\ref{fig:fit} plots the observed 2~p.m.\ temperatures against their fitted values. The relationship is roughly linear and positive; observed and fitted values tend to increase together. Clustering around the smoothed fitted line is somewhat tighter when the fitted values exceed the melting point, with residuals rarely larger than about $\pm2^{\circ}\mathrm{C}$. This results in part from the larger cost for underfitting imposed by the target quantile $\tau = 0.60$. There is greater data sparsity on either side of the temperature range between approximately $-5^{\circ}\mathrm{C}$ and $0^{\circ}\mathrm{C}$. In short, the fit quality produced by quantile gradient boosting is encouraging.\footnote
{
\textcite{Koenker1999} discuss measures of fit for conditional-quantile models derived from the quantile loss function. However, apparently no analogue of the multiple correlation coefficient exists for quantile regression. A suitable measure of fit is introduced later, when conformal prediction regions are presented.
}

Two points merit emphasis. First, the apparent linearity in Figure~\ref{fig:fit} says nothing about whether the lagged predictors themselves relate linearly to temperature. In Figure~\ref{fig:univariate} left panel, the $x$-axis units are days; in Figure~\ref{fig:fit}, they are degrees Celsius. The former shows day-to-day variation in observed temperatures, whereas the latter shows how observed temperatures vary with their fitted counterparts. Second, lagging the predictors means they generate fitted values \emph{two weeks before} the future temperatures are realized. This is not yet true forecasting, because no new unlabeled cases are involved, but it is an important step in that direction.\footnote
{
Fitted values are sometimes called predicted values, which can cause confusion. In this paper, \emph{fitted values} refer to outcomes within the training or calibration data, whereas \emph{forecasted values} refer to predictions for new, unlabeled cases.
}

\begin{figure}[htbp]
\centering
\includegraphics[width=0.75\linewidth]{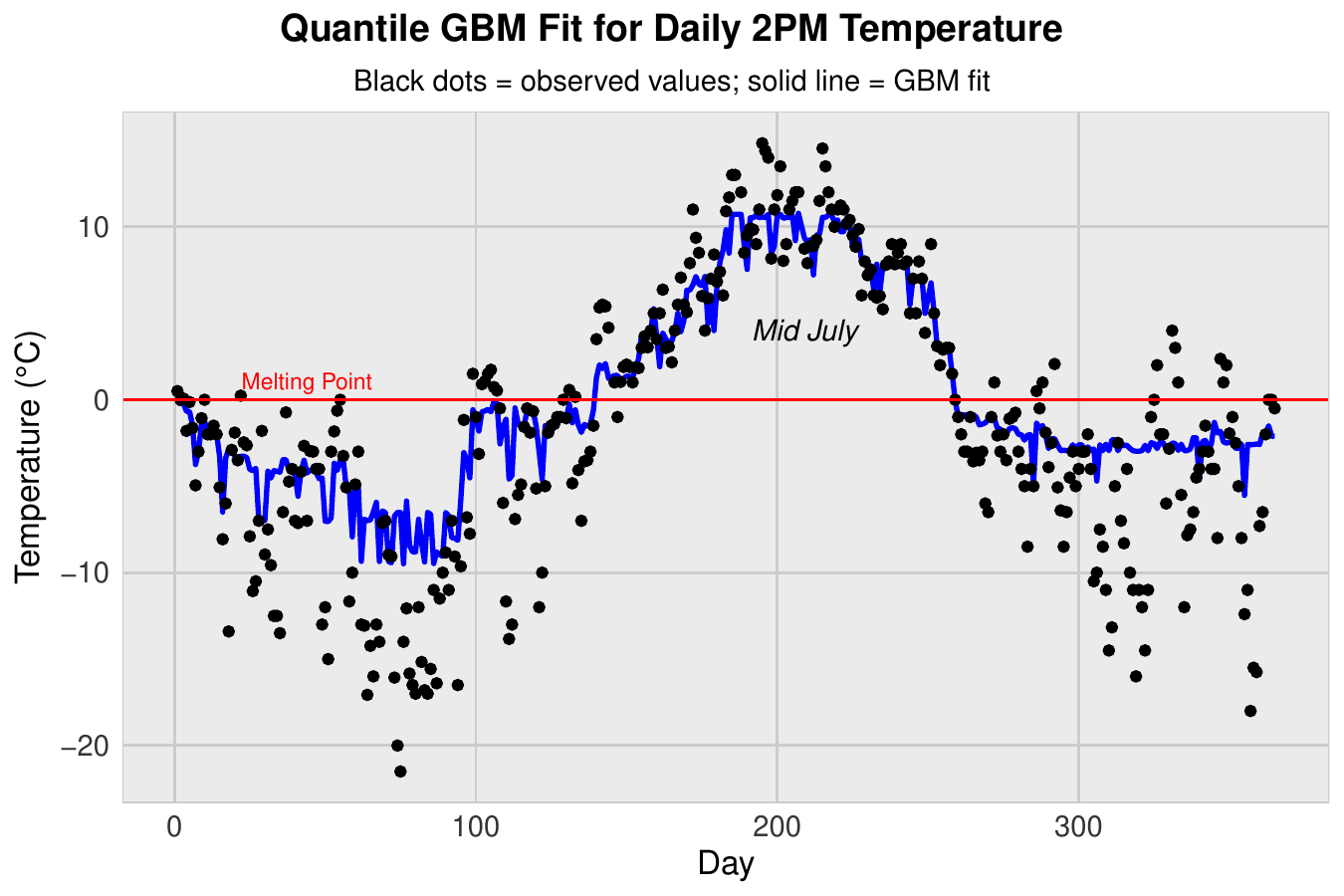}
\caption{Observed and Fitted 2~p.m.\ temperatures from quantile gradient boosting plotted against day of year (2023). Black dots show the observed temperatures over time, the jagged blue line is an interpolation of the fitted values, and the solid red line marks the melting point at $0^{\circ}\mathrm{C}$.}
\label{fig:TS}
\end{figure}

Figure~\ref{fig:TS} provides a complementary perspective that connects more directly to the Arctic-amplification context. It reproduces the observed temperatures and fitted values from Figure~\ref{fig:fit} as time series data, plotted against day of the year. As before, black dots denote the observed 2~p.m.\ temperatures, the jagged blue line in an interpolation of the fitted values, and the horizontal red line marks the melting point.

Overall, the fitted values track the observed temperatures rather closely and capture temperatures above $0^{\circ}\mathrm{C}$ especially well. The unusually warm days clustering in mid-July—around $10^{\circ}\mathrm{C}$—are of particular concern to climate scientists \parencite{Semenov2021}, and they too are fitted accurately, albeit with slight underestimation. Two weeks in advance, the fitted algorithm anticipates melting temperatures effectively, including those mid-July extremes. Higher quantiles such as $Q(0.90)$ would rely on far sparser data and risk instability, making $\tau = 0.60$ a reasonable balance.  As with Figure~\ref{fig:fit}, however, true forecasting remains to be performed.

\subsection{Predictor Impacts on the Fitted Values}

Like all algorithms, quantile gradient boosting is not a model in the traditional statistical sense \parencite{Breiman2001, Kearns2019}. Nevertheless, useful insights about associations in the data can be obtained from variable-importance plots and partial-dependence plots \parencite{Friedman2002}.

\begin{figure}[htbp]
\centering
\includegraphics[width=0.75\linewidth]{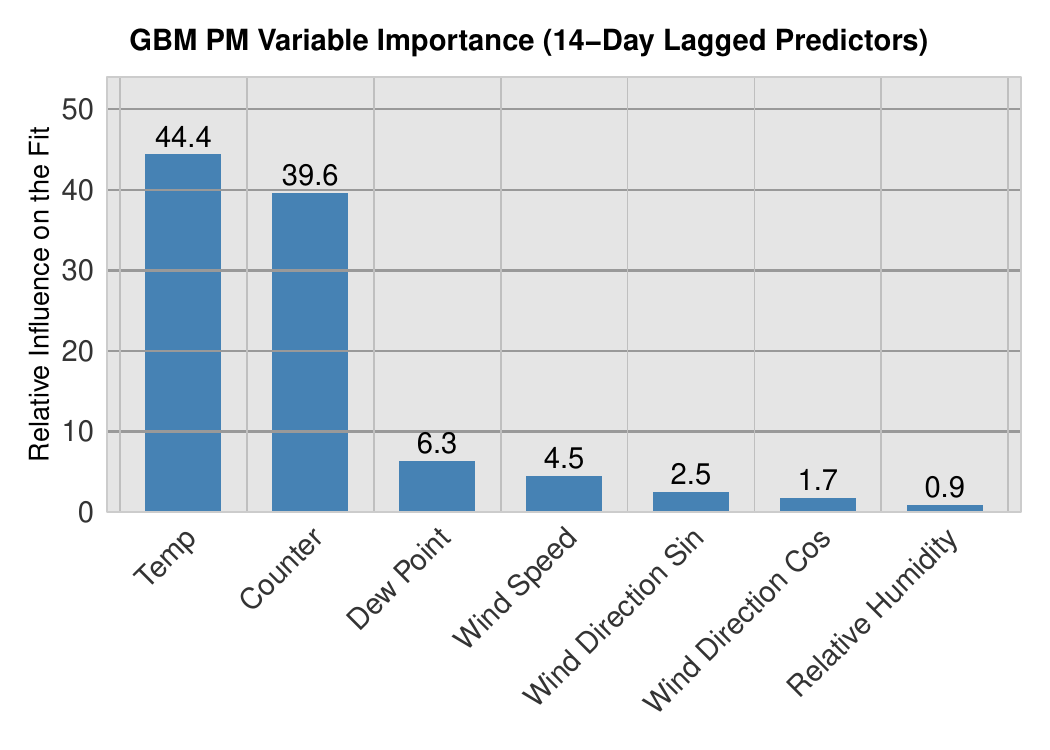}
\caption{Relative contribution of each lagged predictor to the fitted 2~p.m.\ temperatures, computed as the standardized reduction in loss attributable to each predictor.}
\label{fig:imp}
\end{figure}

From the variable-importance plot in Figure~\ref{fig:imp}, the day counter and the 2~p.m.\ temperature lagged by 14~days have by far the strongest associations with the fit of subsequent 2~p.m.\ temperatures. Rounding slightly, the former accounts for about 40\% of the total reduction in loss, and the latter for a bit more than 45\%. The remaining lagged predictors together contribute a little over 10\%. The sum of all contributions is approximately 100\%.

The prominence of the day counter and the 14-day temperature lag is unsurprising. Seasonal patterns are captured by the counter, and the temperature’s gradual evolution over days makes its own lag a strong predictor. The other variables may still capture small or localized temporal effects that, while contributing less to fitting performance, are nonetheless systematically related to the melting temperatures.

\subsection{Functional Forms}

\begin{figure}[htbp]
\centering
\includegraphics[width=0.75\linewidth]{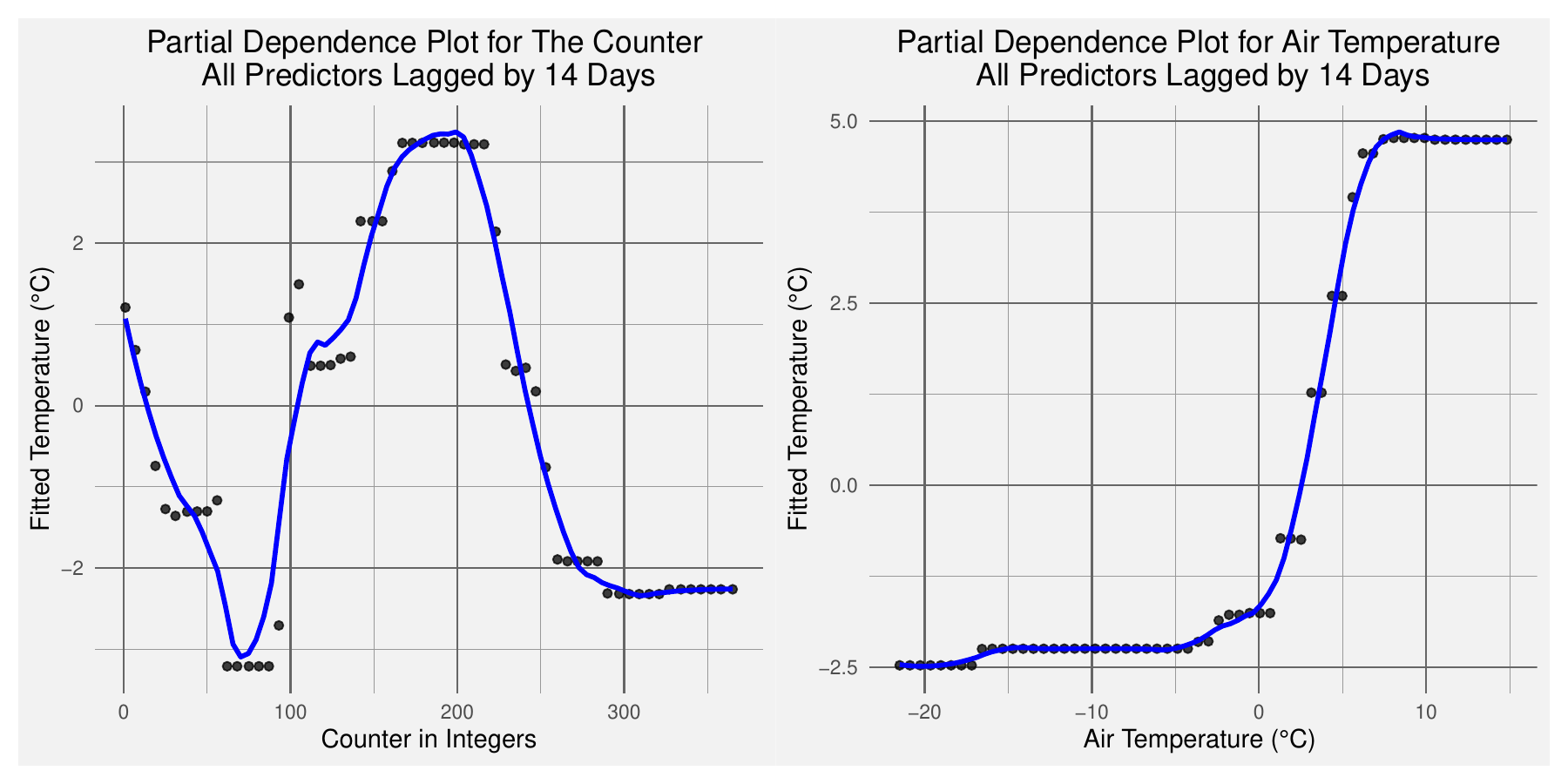}
\caption{Partial-dependence plots for the day counter and the 14-day lagged  2~p.m.\ temperature. The black dots are the partial-dependence values, and the solid blue line is a loess smooth serving as a visual aid. Other predictors distributions are unchanged.}
\label{fig:partial}
\end{figure}

The quantile gradient-boosting algorithm learns associations between each predictor and the response, including the shapes of those relationships. The partial-dependence plots in Figure~\ref{fig:partial} show the relationships between the day counter on the left and the 14-day lagged temperature on the right with the  2~p.m.\ temperature response, each computed with the distributions of all other predictors unchanged.  These are the two variables that dominate the fit. The black dots represent the partial-dependence values, and the solid blue lines show a loess smooth as a visual aid.

Consistent with Figure~\ref{fig:imp}, strong relationships are evident. A look again at Figure~\ref{fig:TS} provides useful context here. The counter-based fitted values range from approximately $-3^{\circ}\mathrm{C}$ to $3^{\circ}\mathrm{C}$, and the lagged-temperature fitted values from about $-2.5^{\circ}\mathrm{C}$ to $5^{\circ}\mathrm{C}$. The first plot is roughly symmetric and concave, peaking during the summer months, consistent with the longitudinal pattern shown in Figure~\ref{fig:TS}. The second plot is S-shaped, with its steepest slope beginning immediately after the melting point at $0^{\circ}\mathrm{C}$ and plateauing near $-2.5^{\circ}\mathrm{C}$ at the left and $5^{\circ}\mathrm{C}$ at the right. The lagged observed temperatures have their strongest association with fitted temperatures just above the melting point. The practical significance of this threshold will be discussed in greater depth below.\footnote
{
The short horizontal strings of dots visible in the plot represent rounding artifacts; the values differ slightly.
}
The remaining four predictors also exhibit nonlinear relationships with the temperature response, but because of their comparatively small contributions to the fit, their partial-dependence plots are omitted in the interest of space.

\section{Forecasts and Estimated Uncertainty}

Addressing uncertainty in a principled way has been anticipated in the research design and algorithmic procedures employed.\footnote
{The material in this section is likely to be unfamiliar to many readers because it differs from traditional efforts to compute forecasting uncertainty. Its recent popularity results from far fewer assumptions that are difficult to test. Somewhat more didactic material is provided.
} 
 The 2024 data serve as a pristine holdout sample of test data that can represent unlabeled cases used in genuine forecasting.\footnote
{
The 2024 data have labels that are useful for didactic purposes and for evaluation of forecasting accuracy. But those labels have no role in the forecasts themselves and would not be known in an operational setting.
}
Forecasts for the 2024 data are produced using the previously fitted quantile gradient-boosting algorithm and its \texttt{predict()} function in \textsf{R}. From the perspective of the trained algorithm, the 2024 lagged predictors constitute new, unlabeled data.

Obtaining the adaptive conformal prediction regions is more involved. All conformal prediction regions begin with a pre-specified coverage probability. By convention, this probability is denoted by $(1-\alpha)$, where $\alpha$ lies between 0 and 0.50. In practice, coverage probabilities range from slightly above 0.50 to nearly 1.0, with values between 0.75 and 0.95 most common. The value of $\alpha$ is determined by subjective considerations shaped by the data, the application and the certainty that subsequent decision makers will likely require. For the conformal approach used here with the Svalbard data, the coverage is $1-\alpha = 0.80$. Other coverage values could be used with reasoning to follow unaffected.

Larger coverage probabilities lead to larger conformal prediction regions. If the goal is to be highly confident that a forecast will be found within a particular prediction region, it makes sense that the length of the prediction region should be long. If the goal is to reduce the length of the prediction region in service of greater precision, confidence that the forecast will be found within that region must decline. Conformal prediction regions formalize this trade-off between coverage and precision. Stakeholders might legitimately prefer a different value for coverage, and within limits imposed by the data, another coverage value is easily implemented.\footnote
{
Ideally, a future decision about a coverage probability would be made with new data. If a new coverage probability is selected for the 2024 data after seeing the results, new complications are introduced and the conformal methods would need some modest changes \parencite{Sarkar2023}.
}

Conformal prediction regions are constructed from ``nonconformal scores.'' These scores can be derived directly from the calibration-data residuals when the calibration data are used as the input into the previously trained boosting algorithm (i.e., \texttt{predict()} in \textsf{R}). Because temperatures from the Svalbard calibration data have strong temporal dependence and the boosting residuals do as well, the residuals were whitened using a first-order autoregressive model (AR(1)). This removes the short-range serial dependence while preserving the level and variability required for valid nonconformal scores.\footnote
{
These steps were guided by estimated autocorrelation functions and Box–Ljung tests determining whether the AR(1) model produced white noise residuals.
}

Exchangeability is required for valid prediction regions \parencite{Vovk2005, Vovk2017, Angelopoulos2025}. One key implication of exchangeability is that it does not matter for the theoretical guarantees whether a test observation is realized after all of the calibration observations are realized. \textcite{Angelopoulos2023} provide an accessible introduction to conformal prediction.

\subsection{Uncertainty Results}

The forecast data from 2024 provide an empirical illustration of how well the forecasting procedures perform. These data have not been used in any way to this point, and as such, share important properties with real forecasting settings that might arise in practice. However, there is additional information, ordinarily unavailable, that can be used to better document forecasting skill: the actual future temperatures being forecast and a full year of observations not just one day at a time.   

For the 2024 test data, the lengths of the conformal prediction regions vary substantially, consistent with Figure~\ref{fig:fit}. For days in which the forecasted 2~p.m.\ temperature exceeds 0, the average half-width is about $\pm 2.6^{\circ}\mathrm{C}$, and the half-widths range from about $1.4^{\circ}\mathrm{C}$ to $6.2^{\circ}\mathrm{C}$. For days in which the forecasted 2~p.m.\ temperature does not exceed 0, the average half-width is about $\pm 4.7^{\circ}\mathrm{C}$, and the half-widths range from about $2.0^{\circ}\mathrm{C}$ to $8.3^{\circ}\mathrm{C}$. Performance is better for the warmer diurnal temperatures because $\tau > 0.50$. 

Stakeholders might not find much comfort in these summary statistics. Even if the mean half-widths promise some guidance for policy and practice, the ranges of the half-widths counsel caution. Perhaps more useful information could be found in estimates of the distributions of the future observed temperatures for each forecasted temperature. In principle, there would be more information, but the sample sizes would be too small, even over several years of data, to support such estimates properly.

In the spirit of conditional distributions of future temperatures, a more modest approach might work. The forecasts can be binned to increase the sample sizes, and rather than estimating full conditional distributions, estimate how likely a future temperature of $0^{\circ}\mathrm{C}$ is exceeded. Treating melting as a “positive”, when a forecasted temperature is above $0^{\circ}\mathrm{C}$ and the future temperature $> 0^{\circ}\mathrm{C}$, a true positive results. When the forecasted temperature is $\le 0^{\circ}\mathrm{C}$ and the future temperature is $> 0^{\circ}\mathrm{C}$, one has a false negative. Ideally, true positives are far more common than false negatives.

\begin{figure}[htbp]
\centering
\includegraphics[width=0.75\linewidth]{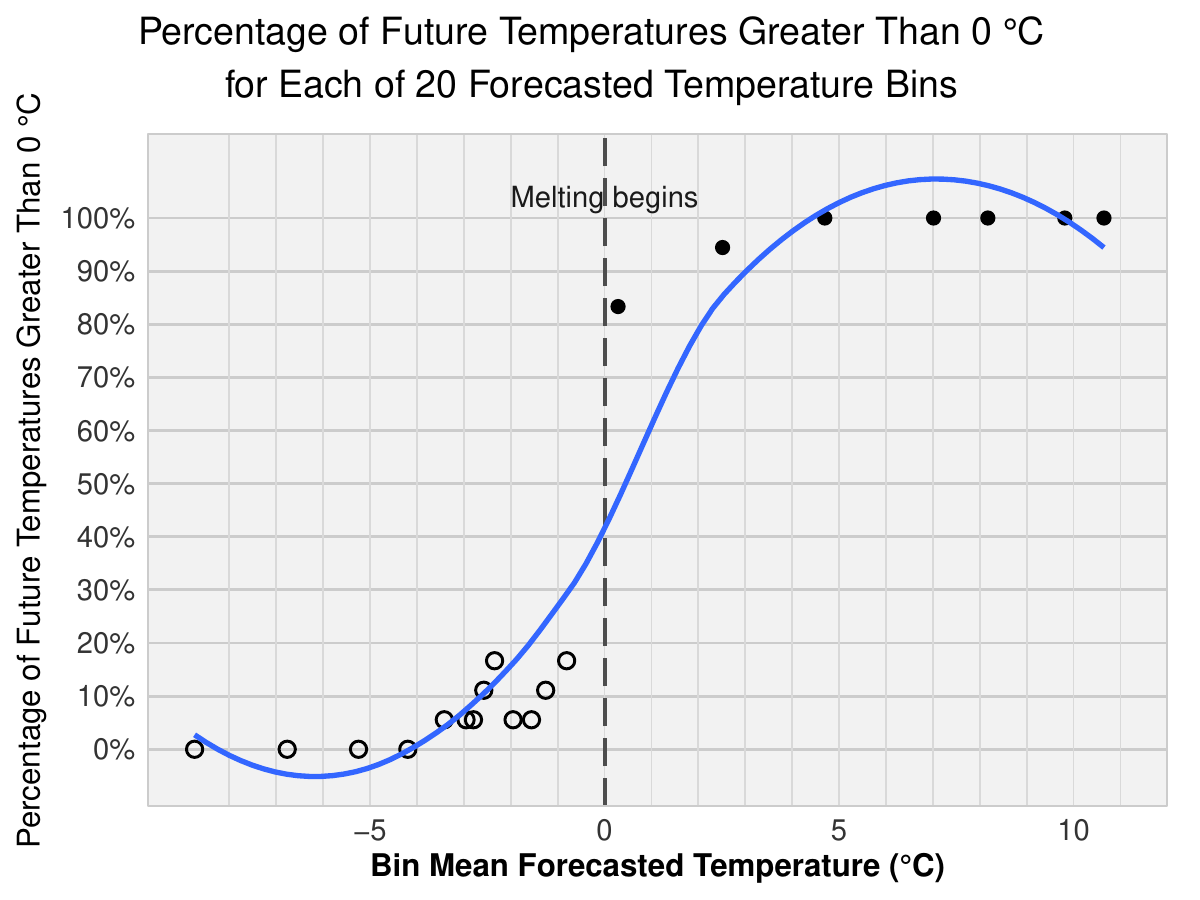}
\caption{For the 2024 holdout forecast data, the mean forecasted temperature in a bin is on the horizontal axis. The percentage of cases in each of 20 equal-width bins having a true future temperature exceeding $0^{\circ}\mathrm{C}$ is on the vertical axis. Filled circles correspond to bins with mean forecasted temperature above $0^{\circ}\mathrm{C}$. Open circles correspond to those below. The solid blue line is a loess smoother.}
\label{fig:prop}
\end{figure}

Figure~\ref{fig:prop} provides a visual summary of these ideas using the 2024 forecast data. The horizontal axis shows the mean of the daily forecasted temperatures for each of 20 equal-width bins. Because the future temperatures are known for each day in 2024 cases, the vertical axis shows the percentage of forecasts within each bin whose true future temperature exceeds $0^{\circ}\mathrm{C}$. (The 14-day lag of each predictor ensures that these are future outcomes.) The loess overlay is S-shaped, with the greatest rate of change between about $-2^{\circ}\mathrm{C}$ and $2^{\circ}\mathrm{C}$.

The conclusions from Figure~\ref{fig:prop} are straightforward. The figure is a confirmation of the analyses above because there is strong evidence of forecasting skill in the holdout sample. As the forecasted temperature increases from about $-10^{\circ}\mathrm{C}$ to about $10^{\circ}\mathrm{C}$, the percentage of cases above the true future melting temperature increases in a nearly monotonic manner from 0.0\% to 100\%. 

Moreover, the percentages on the vertical axis in combination with the smoother confirm that true positives are far more common than false negatives. The largest percentage of false negatives is concentrated between about $-2^{\circ}\mathrm{C}$ and $-1^{\circ}\mathrm{C}$ where a little less than 20\% of the future true temperature values exceed the melting point. As soon as the forecasted temperature exceeds $0^{\circ}\mathrm{C}$, more than 80\% of the forecasts in a bin correspond to future temperatures above $0^{\circ}\mathrm{C}$, reaching 100\% as the bin mean exceeds $4^{\circ}\mathrm{C}$. The large vertical gap between filled and open circles highlights the relatively abrupt phase transition centered on $0^{\circ}\mathrm{C}$ responsible for the dramatic increases in true positives. Because melting is such an important event, stakeholders might find Figure~\ref{fig:prop} encouraging. Some kinds of temperature forecasts might be accurate enough to be useful.

The loess curve in Figure~\ref{fig:prop} shows a gradual transition around  $0^{\circ}\mathrm{C}$ rather than an abrupt step function. The physics of melting helps explain why. Melting is a complicated process \citep{Noto2026} and can vary for different kinds of ice \citep{Wadhams2000}.  For example, when sea ice forms, its hexagonal lattice cannot readily incorporate salt ions. The excluded salt forms brine pockets and channels among the ice grains, leaving the ice as nearly fresh water while the brine becomes increasingly salty. As freezing continues, the concentrated brine's freezing point can fall well below $0^{\circ}\mathrm{C}$ because the salt ions attract water molecules the lattice needs; it is more difficult for the water molecules to self-organize into the lattice structure. Unlike fresh water ice, which melts at a single fixed temperature, sea ice melts over a range of temperatures, because the brine pockets within it are already liquid at temperatures below $0^{\circ}\mathrm{C}$. As the temperature rises, progressively more of the remaining ice melts, and the brine channels widen and connect. Because a significant fraction of sea ice consists of these pre-existing liquid brine inclusions, it requires less energy to melt than an equivalent mass of fresh water ice — the brine pockets need no latent heat to liquefy, only the nearly pure ice fraction does. By the same logic, the energy released when salt water freezes is less than that released when fresh water freezes. This interpretation aligns with Figure~\ref{fig:prop}.

From a policy perspective, even approximate information about when melting will occur can be valuable. Although graphs like Figure~\ref{fig:prop} cannot be produced prospectively—since future temperatures are unknown—they can be constructed retrospectively using historical weather-station data. Such analyses help stakeholders gauge likely rates of true positives and false negatives conditional on forecasted temperature. For example, if a forecast exceeds $4^{\circ}\mathrm{C}$, the realized future temperature will almost certainly be $> 0^{\circ}\mathrm{C}$.\footnote
{
The plotted percentages should not be interpreted as probabilities. Temporal dependence persists among the forecasted temperatures within each bin, although its structure is not easily characterized. The forecasts are ordered by temperature rather than by time, and the time gaps between forecasts within a bin can range from days to months. For example, forecasted temperatures during late spring and early fall may be similar even though they occur several months apart.
}

Given these results, the role of adaptive conformal prediction regions can be briefly revisited. For a new, unlabeled case $T{+}1$, the true temperature will fall within its adaptive conformal prediction region with probability at least $1-\alpha$. When the lower bound of a prediction region is greater than zero, the probability that the future temperature will be above 0 is at least $1-\alpha/2$. For a specified coverage probability of 0.80, that probability would be at least 0.90. Coverage is gained because under some circumstances, the prediction region's upper bound may not matter much. This approach could work well for Svalbard insofar as temperatures $> 0^{\circ}\mathrm{C}$ are far more concerning than lower temperatures. In the spirit of Figure~\ref{fig:prop}, when the lower bound of the prediction region is greater than $> 0^{\circ}\mathrm{C}$, one may simply assume that the forecast of melting is correct.

\section{Discussion}

The analysis of 2~p.m.\ temperatures in Svalbard, Norway, combines three interlocking statistical traditions: gradient boosting with a quantile as the estimation target, an AR(1) model to construct exchangeable nonconformal scores, and adaptive conformal prediction regions computed with quantile random forests, a method robust to sparse data." To the best of the author’s knowledge, these procedures applied to three temporally separated samples has not previously appeared in the literature.

A two-week forecasting lead time before the onset of widespread surface melting is a narrow window, but it would give Arctic community administrators critical time to prepare for infrastructure resilience, public safety, and logistical challenges \parencite{Streletskiy2019,USARC2003}. Local governments could limit heavy-vehicle use on thaw-sensitive roads, runways, and around pipelines; pre-position maintenance materials; and inspect culverts, bridges, and drainage channels before damage occurs \parencite{ARCUS2021}. Water managers could adjust reservoir levels or activate temporary treatment measures to mitigate siltation and contamination from meltwater inflows.

In communities dependent on ice roads, snowmobile trails, or frozen river crossings, early warnings would permit orderly resupply and fuel delivery before surface travel becomes unsafe \parencite{Chen2025}. Coastal and hillside settlements could prepare for increased risks of shoreline erosion or permafrost-related landslides \parencite{Firelight2022}. Health authorities might use the lead time to issue advisories on water quality or vector-borne disease risk as standing water accumulates.\footnote
{ 
Because of Arctic warming, even Iceland is now reported to host mosquitoes \parencite{Straker2025}.
}

More broadly, forecasts on a two-week timescale could foster coordination among local administrators, regional governments, and research stations, improving situational awareness and resource allocation across sectors. Even modest improvements in lead time may yield substantial adaptive benefits in Arctic regions where melt conditions evolve rapidly and logistical flexibility is limited.

\section{Conclusions}

Projections of future global warming have long been central to IPCC assessments. The analyses presented here contribute to the complementary literature on smaller-scale local forecasting. Temperature forecasts with a two-week lead time show promise, especially when focused on Arctic thresholds associated with widespread melting. Forecasting uncertainty is explicitly addressed with adaptive conformal prediction regions, and results are presented in a manner intended to be accessible and useful to stakeholders. With thoughtfully applied algorithms and accessible data, informative short-term forecasts can be generated on an ordinary desktop computer with an internet connection.

\section*{Appendix: Statistical Narrative}

The adaptive conformal forecasting procedure proceeds sequentially in the following steps. Separate calibration and forecasting steps ensure that all prediction intervals
are based solely on information available at the time of forecasting. The sequence below is written with \textsf{R} in mind but also applies to Python.

\begin{enumerate}

\item \textbf{Data preparation.}  
      Subset the longitudinal Longyearbyen weather-station data into training data from 2023, calibration data from 2022, and test data from 2024.  Specify a coverage probability $1-\alpha$.  For each year, the response $y_t$ is the daily 2~p.m.\ air temperature in degrees Celsius.  The predictors for each dataset $X_{t-14}$ are the same meteorological variables lagged by 14 days.

\item \textbf{Algorithm training.}  
      With $\tau = 0.60$, fit quantile gradient boosting to the 2023 training data 
      (\texttt{train}), yielding a trained algorithm (\texttt{gbm}) and an optimal iteration count (\texttt{best.iter}).

\item \textbf{Residuals for calibration.}  
      Using the 2022 calibration data (\texttt{calibration}), compute fitted values
      \[
      \widehat{y}_t^{2022} 
        = \texttt{predict}(\texttt{gbm}, \texttt{calibration}, \texttt{best.iter}),
      \]
      and residuals 
      \[
         r_t^{2022} = y_t^{2022} - \widehat{y}_t^{2022}.
      \]

\item \textbf{Regime split.}  
      In response to subject-matter concerns, split the calibration residuals into warm and cool regimes using the fitted values:
      \[
      \text{Warm if } \widehat{y}_t^{2022} > 0, 
      \qquad 
      \text{Cool if } \widehat{y}_t^{2022} \le 0.
      \]

\item \textbf{Temporal-dependence adjustment.}  
      Because of temporal dependence in $ r_t^{2022}$, fit a first-order autoregressive model with intercept to the \emph{full} residual sequence,
      \[
      r_t^{2022} = \mu + \phi r_{t-1}^{2022} + \varepsilon_t.
      \]
      This produces a single set of AR(1) innovations. These innovations are then allocated to the warm and cool regimes according to the fitted-values split, which removes short-range serial dependence without altering the level or variability.

\item \textbf{Innovations (nonconformal scores).}  
      Extract the AR(1) innovations $\varepsilon_t$ for each regime. These approximately white-noise innovations serve as the nonconformal scores.

\item \textbf{Quantile-function estimation.}
      For each regime, fit a quantile random forest (QRF) to the 2022 nonconformal 
      scores, using the 2022 fitted values and the 2022 lagged predictors as covariates.  
      QRF learns how the variability of the nonconformal scores depends on the 
      fitted temperature level and on the lagged meteorological predictors.
      
\item \textbf{Forecast generation.}  
      Using the 2024 test data (\texttt{forecast}), compute \texttt{gbm} forecasts
      \[
      \widehat{y}_t^{2024} 
         = \texttt{predict}(\texttt{gbm}, \texttt{forecast}, \texttt{best.iter}),
      \]
      and assign each case to the warm or cool regime according to the sign of its forecast.

\item \textbf{Nonconformal-score quantiles.}
      For each 2024 case, supply its forecasted value and lagged predictors to 
      the corresponding regime’s QRF to obtain predicted score quantiles 
      $\widehat{q}_{\alpha/2}$ and $\widehat{q}_{1-\alpha/2}$.

\item \textbf{Prediction intervals.}  
      Construct lower and upper adaptive conformal prediction bounds:
      \[
      \mathrm{PI}_{\mathrm{lower}}
         = \widehat{y}_t^{2024} + \widehat{q}_{\alpha/2},
      \qquad
      \mathrm{PI}_{\mathrm{upper}}
         = \widehat{y}_t^{2024} + \widehat{q}_{1-\alpha/2}.
      \]
      The interval widths reflect both the variability captured by QRF and any remaining 
      dispersion in the nonconformal scores.
\end{enumerate}

These linked procedures combine quantile gradient boosting, AR(1) whitening, and quantile-random-forest calibration to yield adaptive conformal prediction regions.  The lower and upper bounds use the ranks of the nonconformal scores. 

\subsection*{Exchangeability of Nonconformal Scores}

The supervised learning algorithm is first fitted on the training data  
$\{(X_t, Y_t)\}_{t=1}^{T_{\mathrm{train}}}$. Predictions are then generated for the separate calibration data $\{X_t\}_{t=1}^{T}$, and their AR(1) innovations form the nonconformal scores $\{S_t\}_{t=1}^{T}$.  

If $\{S_t\}_{t=1}^{T+1}$ are (approximately) exchangeable then, conditional on the fitted algorithm, the calibration scores and the new score
$S_{T+1}$ are also exchangeable.  
Consequently,
\[
\Pr\!\left\{\,S_{T+1} \le 
     q_{1-\alpha}\big(\{S_t\}_{t=1}^{T} \cup \{S_{T+1}\}\big)\right\}
     \ge 1-\alpha,
\]
and the adaptive conformal prediction region attains marginal coverage of at least $1-\alpha$.

Exchangeability means that the joint distribution of the nonconformal scores is invariant to permutations of their indices.  Although in practice the scores are realized in temporal order, their joint law would be unchanged had they been realized in any other order.  That a new unlabeled case is realized at time $T{+}1$ is immaterial.  
For the analyses of the Svalbard data, exchangeability depends on the whitening of the calibration data residuals. After whitening, the resulting innovations satisfy the exchangeability condition sufficiently well to justify the coverage guarantee above.

\bibliographystyle{imsart-nameyear} 
\bibliography{IceMelt}

\end{document}